\documentclass[a4paper,11pt]{article}
\usepackage[T1]{fontenc}
\usepackage[utf8]{inputenc}
\usepackage{graphicx}
\usepackage{multicol}

\usepackage{hyperref}
\hypersetup{
  setpagesize  = false,
  colorlinks   = true,    
  urlcolor     = blue,    
  linkcolor    = black,   
  citecolor    = black    
}
\usepackage[outercaption]{sidecap}

\usepackage{authblk}
\usepackage{geometry}
\usepackage{setspace}
\usepackage{verbatim}
\usepackage[utf8]{inputenc}
\usepackage{amsmath}
\usepackage{amssymb}
\usepackage{gensymb}
\usepackage{verbatim} 
\usepackage{amsfonts}
\usepackage{cancel}

\usepackage{marginnote}

\usepackage[sorting=none]{biblatex}
\newcommand{\keywordsenglishname}{Keywords}

\renewenvironment{abstract}{%
        \begin{center}
	\begin{minipage}{14cm}
	{\textbf{\abstractname:}}
}{
        \end{minipage}
	\end{center}
}
\newenvironment{abstractinenglish}{
        \def\abstractname{\abstractinenglishname}
	\begin{abstract}
}{
        \end{abstract}
}

\newenvironment{keywordsenglish}{
        \def\abstractname{\emph{\keywordsenglishname}}
	\begin{abstract}
}{
        \end{abstract}
}

\title {Experimental Investigation of Electromagnetically Induced Transparency in Selective Reflection Spectra}
\author{Armen Sargsyan, Anahit Gogyan$^{\dagger}$, David Sarkisyan}
\affil{Institute for Physical Research, National Academy of Sciences of Armenia, Gitavan-2, 0204 Ashtarak, Armenia}
\affil{\small{$^\dagger$ Corr. author: agogyan@gmail.com}}
\date{}
\usepackage{fancyhdr}
\begin{document}

\maketitle
\vspace{6pt}

\begin{abstractinenglish}
We have investigated electromagnetically induced transparency in the spectrum of selective reflection at the interface of Rb atom vapors and a dielectric nanocell window. A nanocell with atomic vapor column thicknesses ranging from 150 to 1200 nm, as well as a 50~$\mu$m thickness microcell were used. 
We have compared electromagnetically induced transparency observed for the cases of the selective reflection and transmission. It was demonstrated that for the thicknesses of below $\leq 1000$~nm selective reflection technique is more favorable. In contrast, for wider cells and microcells, using transmitted radiation as probe field is more effective.
\end{abstractinenglish}

\begin{keywordsenglish}
 electromagnetically induced transparency;  selective reflection;  nanometric thin cell   
\end{keywordsenglish}
 
\vspace{6pt}
\section{Introduction}\label{intro}

Electromagnetically induced transparency (EIT) is a quantum interference phenomenon where an otherwise opaque medium becomes transparent to a probe laser beam when a control laser beam is applied \cite{13}. This effect occurs due to the coherent superposition of atomic states, leading to the cancellation of absorption for the probe light. EIT is used in various applications, including slow light propagation, quantum memory, and enhanced nonlinear optical processes \cite{13,14,15}. Selective reflection (SR), on the other hand, involves the preferential reflection of specific wavelengths of light at the interface between two media, typically due to resonant interactions with atomic or molecular transitions \cite{1,2,3,4,5}. SR is employed to probe and analyze the properties of atomic vapors, providing valuable insights into phenomena such as EIT and other atomic-level interactions. Particularly, in \cite{12}, the theoretical study has been conducted on EIT process examining the role of SR and the sensitivity of the spectra to the optical length was demonstrated.

Interest in SR stems from several significant factors. Firstly, SR possesses a sub-Doppler spectrum width and high reflectance, making it highly advantageous for detailed spectral analysis. Secondly, SR facilitates the examination of spectrum broadenings caused by atomic collisions at high densities and the identification of frequency shifts in atomic transitions. Thirdly, SR has been successfully employed to study atom-wall interactions, allowing for the measurement of van der Waals forces. This is evidenced by a red shift in the SR frequency at atomic distances from windows less than 100~nm \cite{4,6,7,8,9}. Additionally, the dispersive shape of the SR spectrum can be used to stabilize the frequency of continuous wave diode lasers, thereby enhancing their performance \cite{10}. Finally, the influence of magnetic fields ranging from 0.1 to 6~kG on the SR spectrum can been studied using a nanocell containing atomic vapor of Cs, Rb, and $^{39}$K, demonstrating the feasibility of remotely determining the magnitude of magnetic fields \cite{5,7,10,11}.
\begin{figure}[b]
    \centering
    \includegraphics[width=0.48\textwidth]{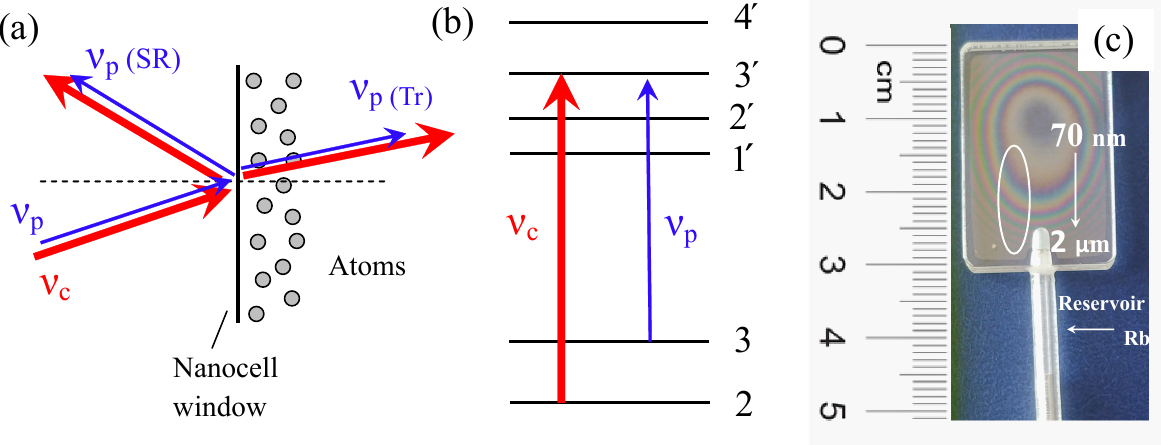}
    \caption{(a) Fragment of the experimental scheme for EIT formation in a nanocell containing Rb atomic vapor. Blue arrows are paths of probe light, which is partially reflected from atomic vapor and partially passing through the nanocell. Red arrows are coupling field. (b) Diagram of the $\Lambda$-system formed on the $^{85}$Rb $D_2$ line. (c) Photograph of the nanocell. Interference fringes, formed by the reflection of light from the inner surfaces of the windows, are visible. The oval marks the gap region, which ranges from 70~nm to 2~$\mu$m. The lower part shows a sapphire tube, which serves as a side-arm reservoir filled with Rb metal.}
    \label{fig:Fig1}
\end{figure}
In this work, the EIT process was experimentally studied for the first time with the involvement of selective reflection from the interface between Rb atomic vapors and the dielectric (sapphire) window of the nanocell. The vapor column of atoms enclosed in the nanocell had a thickness varying between 150 and 2000 nm, and a separate cell with a vapor column thickness of approximately 50~$\mu$m was also utilized. A comparison was made between EIT resonances formed when SR was used as the probe radiation, and when the probe radiation transmission through the nanocell was used.

In the next section we present the experimental details and obtained results with discussions, results are concluded in the last section.

\section{Experimental Procedures and Results}\label{exp}

\subsection{Nanocell of a variable thickness}

We have studied EIT experimentally in Rb atomic vapor $D_2$ line. We have used two diode lasers with a wavelength of 780~nm and a linewidth of less than 1~MHz \cite{16}. Schematics of the light interaction with atoms is partially shown in Fig 1a. SR occurs at the interface between the Rb atomic vapors and the sapphire window of the nanocell \cite{4,5}, where reflected light plays the role of the probe radiation of frequency $\nu_P$. The second sapphire window of the nanocell, positioned very close to the first window (not shown to avoid overloading the figure), has an internal surface nearly parallel to that of the first window, making the nanocell function like a low-finesse Fabry-Perot etalon \cite{5,10,17}. The distance between the inner surfaces of the windows, referred to as the gap size $L$, can be smoothly varied within the range of 30-2000~nm by ensuring a slight wedge shape to the gap. The method for determining the nanocell gap thickness is detailed in \cite{17}. When the gap thickness $L = n\lambda/2$, where $n$ is an integer, due to the destructive interference SR power gets zero. Conversely, at gap thicknesses of $L = (2n+1)\lambda/4$, including $n=0$, constructive interference causes the SR power to reach up to approximately 20\% of the radiation power directed to the nanocell \cite{5,7,17}.

A picture of the nanocell filled with Rb atomic vapor is shown in Fig. 1c, with detailed fabrication information provided in \cite{18}. The nanocell is positioned in a two-section heater, allowing independent heating of the windows and the side-arm reservoir containing Rb metal. The heater features two openings, each with a diameter of 10~mm, to facilitate the passage of laser light. By smoothly adjusting the vertical position of the heater with the nanocell inside, laser radiation can traverse through varying thicknesses of the nanocell.

The probe radiation of frequency $\nu_P$ and the coupling radiation of frequency $\nu_C$ are combined and directed perpendicularly to the window of the nanocell window filled with Rb atomic vapor. The frequency of the probe radiation $\nu_P$ is scanned along the $F_g = 3 \rightarrow F_e = 2', 3', 4'$ transitions, while the coupling radiation frequency is in resonance with the $F_g = 2 \rightarrow F_e = 3'$ transition, forming a $\Lambda$-system shown in Fig. 1b. A portion of the probe light $\nu_P$ is directed to an auxiliary Rb atomic vapor nanocell with a thickness $L = \lambda = 780$~nm, which acts as a reference transmission spectrum \cite{19,20}. The powers of the coupling and probe lights of linear and mutually perpendicular polarizations are 2.5~mW and 0.1~mW, respectively. This configuration allows to transmit only the probe field when using a polarizer and record its spectrum, while blocking the coupling field \cite{20}. Atomic vapor temperature is approximately 100$^\circ$C, corresponding to an atomic vapor density of $N \approx 6 \times 10^{12} \, \text{cm}^{-3}$. We register the EIT signal in the SR light, denoted as EIT$_{SR}$ below and compare with the EIT in the transmitted light, labeled as EIT$_{T}$. 

\sidecaptionvpos{figure}{c}
\begin{SCfigure*}
   \includegraphics[width=0.48\textwidth]{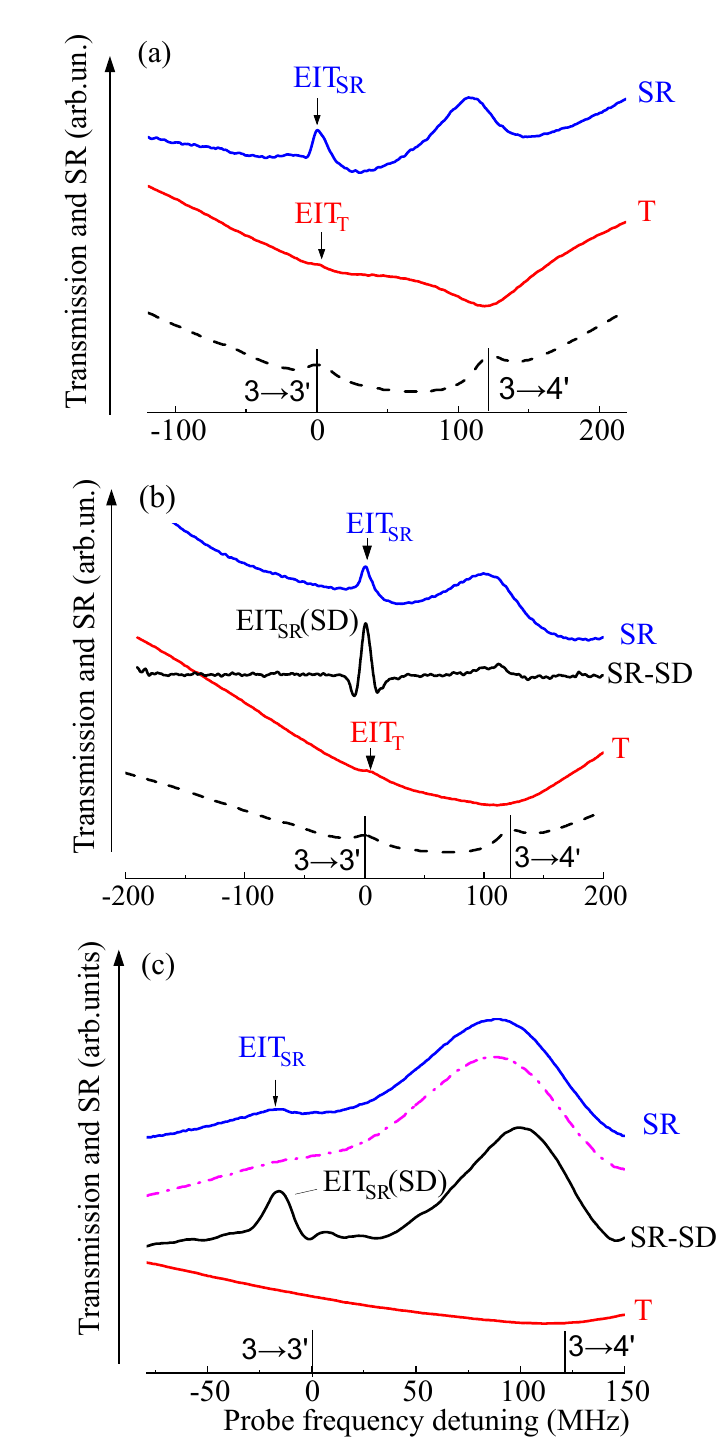}
     \caption{Spectra of $^{85}$Rb atoms D$_2$ line for different widths of atomic vapors. \textit{Blue, SR labeled}curves are the back reflected SR radiation, \textit{red, T labeled} curves are the spectrum of the probe radiation recorded in the forward direction, dashed lines represent the reference signal. EIT$_{SR}$ and EIT$_{T}$ are labeled. 
    (a) Nanocell thickness of 1100~nm. 
    (b) Nanocell thickness of 700~nm. Black solid line is the second derivative of the blue line. 
    (c) Nanocell thickness of 150~nm. Magenta dash-dotted curve is back reflected light when no coupling field is applied. Black solid line, labeled as SR-SD,  is the second derivative of the blue line.
    Transitions $F_g = 3 \rightarrow  3' \text{and} \ 4'$ are labeled as frequency references.}
\end{SCfigure*}

Recorded back reflected and transmitted light intensities dependence on probe light detuning are presented in Fig. 2a, 2b and 2c for nanocell thicknesses of $L \sim 1100$~nm, $L \sim 700$~nm and $L \sim 150$~nm, resepctively. A well-known fact about SR formation is that it occurs over lengths on the order of the evanescent fields, approximately $\lambda/2\pi$ \cite{4,7}. This is why the cell lengths are chosen accordingly.

In all figures, the blue curve labeled "SR" represents the back-reflected SR radiation dispersive spectrum, while the red curve labeled "T" depicts the spectrum of the probe radiation recorded in the forward direction after transmission through the atomic vapors.
Dashed lines represent the reference signal with labeled  transitions $F_g = 3 \rightarrow  3' \text{and} \ 4'$. Observed EIT peaks are labeled as EIT$_{SR}$ and EIT$_{T}$ for the selectively reflected and transmitted radiations.

In Fig. 2a, the atomic vapor column is approximately 1100 nm, and it is evident that the EIT$_{T}$ is weakly manifested compared to EIT$_{SR}$, which is more pronounced. Additionally, an increase in the absorption of probe radiation at the transition $F_g = 3 \rightarrow F_e = 4'$ is recorded in the transmitted light, due to an increased population of Rb atoms at the ground state $F_g = 3$ as a result of optical pumping by the coupling radiation \cite{21}. 

In Fig. 2b, the SR curve contains a well-resolved EIT$_{SR}$ resonance. The nanocell gap thickness is approximately $L \sim 700$~nm. As demonstrated in \cite{22}, when a spectrum includes resonances of varying widths, applying double differentiation, enhances the amplitudes of narrow-width resonances. This is particularly useful when the spectrum contains an EIT-resonance. The black curve labeled "SR-SD" represents the second derivative (SD) of the SR spectrum. Since the spectral width of the EIT$_{SR}$ resonance is expected to be narrower than that of the SR, the amplitude difference becomes more pronounced in the second derivative spectrum. The transmission curve T contains the EIT$_T$ resonance, which is only weakly observed. 


\begin{figure}
    \centering
    \includegraphics[width=0.7\textwidth]{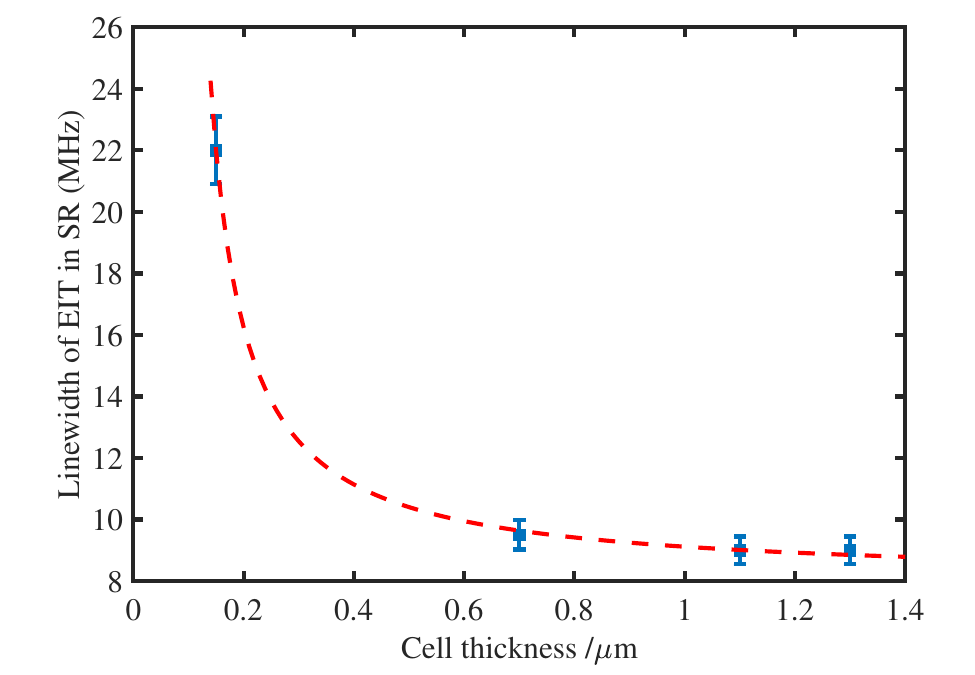}
    \caption{The spectral width of the EIT$_{SR}$ resonance increases as the thickness $L$ of the Rb vapor column decreases from 1.5 µm to 0.15 µm. This broadening is attributed to collisions of the Rb atoms with the nanocell windows.  A fitted reciprocal curve is shown by red dashed line. 
    }
    \label{fig:Fig3}
\end{figure}

In Fig. 2c, for a nanocell thickness of approximately 150 nm, a weakly spectrally resolved EIT$_{SR}$ resonance is formed. For comparison, the probe field SR spectrum when the coupling radiation is off is also shown by the magenta dash-dotted line in Fig. 2c. The second derivative of SR spectrum, shown by black solid line, significantly improves the spectral resolution of the EIT$_{SR}$ resonance.
Probe field transmission spectrum recorded in the forward direction in red line shows the absence of the EIT$_{T}$ resonance. Thus, the EIT resonance is still detectable in the SR spectrum, whereas no EIT is detected in the transmission spectrum.

Note that the results presented above were obtained when the laser frequency was in resonance with the corresponding atomic transition. In this scenario, the EIT process involves atoms flying nearly parallel to the nanocell’s windows, resulting in a relatively low number of collisions between the atoms and the windows. However, the situation changes significantly when there is a frequency mismatch, detuning $\Delta$, between the frequencies of the coupling radiation and the corresponding atomic transition \cite{20}. In this case, the formation of EIT involves atoms traveling parallel to the laser radiation, along the $z$-axis, for which collisions with the nanocell’s windows become significant.  It is known that a unique collision with a dielectric surface of an uncoated vapor cell is suffcient to thermalize the ground hyperfine levels, with depolarization probability 0.5 - 1 \cite{23nn}.


As the thickness of the vapor cell is reduced, the lifetime of the ground-state coherence, $\gamma_{c}^{-1}$ becomes shorter due to collisions of atoms with the cell windows travelling in parallel to the laser radiation, along the $z$-axis. The relationship is given by 
\begin{equation}
    \gamma_{c} = \frac{1}{2\pi \tau_L},  \label{eq:eq1}  
\end{equation}
where $\tau_L = \frac{L}{v_z}$ is the flight time with $L$ being the distance between the inner surfaces of the windows and $v_z = \frac{2\pi\Delta}{k} = \lambda \Delta $ the projection of the most probable thermal velocity on $z$ axis. $k$ and $\lambda$ are atomic number and wavelength, respectively. Equation \eqref{eq:eq1} can be re-written as
\begin{equation}
    \gamma_{c} = \frac{\lambda \Delta }{2\pi} \frac{1}{L}. \label{eq:eq2}
\end{equation}
The significant influence of detuning $\Delta$ on the deterioration of contrast and broadening of the EIT-resonance spectrum was studied in detail in Ref. \cite{20}.


A decrease in the atomic vapor thickness $L$ leads to an increase in the spectral width of the SR spectrum, causing a rapid broadening of the EIT$_{SR}$ resonance. This effect is illustrated in Fig. 3, which shows the thickness decreasing from 1.5 $\mu$m to 0.15 $\mu$m. 

As indicated by our previous results, when the detuning $\Delta \approx 0$, the width of the optical resonances (SR and EIT) is inversely proportional to the thickness of the atomic vapor column \cite{7,23n}. Consequently, the dotted red curve in Fig. 3 is fitted by an inverse function $\sim 1/L$.




\subsection{Microcell of thickness of 50~$\mu$m}

In Ref. \cite{19} it was demonstrated that velocity-selective optical pumping (VSOP) resonances appear in the absorption spectrum of a nanocell with $L = \lambda$ thickness when the laser radiation intensity exceeds 1~mW/cm², resulting in reduced absorption. These VSOP resonances at the atomic transitions have a spectral width approximately ten times narrower than the Doppler width, making them useful as frequency markers.

For the formation of VSOP resonances, atoms with a velocity projection $v_z$ contribute. However, contribution of these atoms quickly disappear due to collisions with the nanocell windows. Thus, VSOP resonances, that are located at the atomic transitions, are predominantly formed by atoms flying parallel to the nanocell windows with $v_z = 0$. This is one advantage of using a nanocell for EIT formation. In a long cell, the presence of probe and coupling radiation leads to the formation of many VSOP resonances, making the spectrum complex \cite{23}. For the $D_2$ line, the number of VSOP resonances is greater than for the $D_1$ line because there are more upper levels in the hyperfine structure, and the frequency spacing between them is smaller.

To investigate the dependence of the ratio of the amplitudes of the EIT$_{SR}$ and the EIT$_{T}$ resonances on the thickness $L$, a cell with a thickness of $L = 50\ \mu\text{m}$, filled with Rb atomic vapor, was studied. In this setup, the frequency of the probe radiation is scanned along the transitions $F_g=2 \rightarrow 2', 3'$, while the coupling frequency is in resonance with the $F_g=3 \rightarrow 2'$ transition to form the $\Lambda$-system. The powers of the coupling and probe beams are 5 mW and 50 $\mu$W, respectively.
\begin{figure}[bh!]
    \centering
    \includegraphics[width=0.48\textwidth]{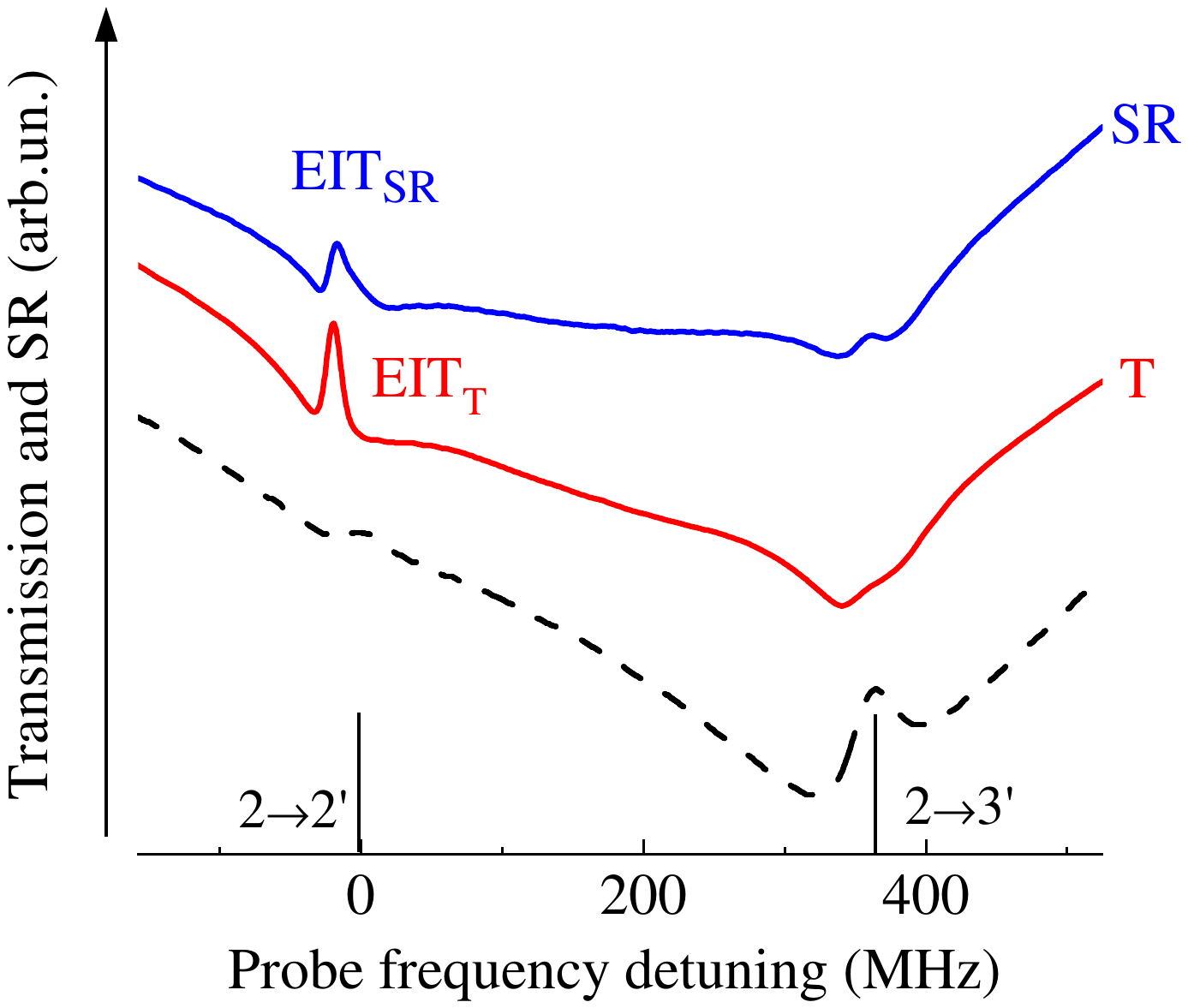}
    \caption{Spectra for a microcell with a thickness $L = 50~\mu$m filled with Rb vapor, $D_1$ line. Blue line is the SR spectrum of the probe radiation recorded in the backward direction, which contains the EIT$_{SR}$ resonance. Red line is the transmission spectrum of the probe radiation recorded in the forward direction, where the EIT$_{T}$ resonance is manifested with a 15\% contrast. Black dashed line is the reference spectrum with labeled $F_g = 2 \rightarrow F_e' = 2', 3'$ transitions.}
    \label{fig:Fig4}
\end{figure}

In Fig. 4, blue, SR labeled curve shows the spectrum of the SR probe light recorded in the backward direction, containing the EIT$_{SR}$ resonance. Red, T labeled curve shows the spectrum of the probe radiation recorded in the forward direction, after transmission through the atomic vapors, where a pronounced EIT$_{T}$ resonance is manifested. The contrast of the EIT$_{T}$ resonance, defined as the ratio of the EIT$_{T}$ resonance amplitude to the peak absorption of atomic vapors in the absence of the coupling radiation, is 15\%  at a microcell temperature of $70\ ^\circ\text{C}$, which corresponds to the density of $N \approx 7.4 \times 10^{11}\ \text{cm}^{-3}$. At this relatively low vapor density, the SR process is not strong. However, an increase in temperature (and hence the absorption of atomic vapors) will lead to a decrease in the magnitude of the EIT$_{T}$ resonance contrast.

An increase in the absorption of probe radiation at the $F_g=2 \rightarrow 3'$ transition is also observed in the spectrum, due to an increase in the population at the $F_g=2$ level as a result of optical pumping by the coupling radiation. The full spectral width of the EIT$_{T}$ resonance at half maximum is approximately 9~MHz {for 50~$\mu$m nanocell and for $D_1$ line}. As seen, the formed EIT$_{T}$ resonance exhibits better parameters (amplitude and spectral width) compared to the EIT$_{SR}$ resonance.


The power-broadened EIT-resonance linewidth can be calculated using the expression given in \cite{15}:
\begin{equation}
\gamma_{\text{EIT}} = \frac{\Omega_C^2}{\Gamma_{\text{Dop}}} + \gamma_{c},\label{eq:eq3}
\end{equation}
where $\Omega_C$ is the Rabi frequency of the coupling radiation, and $\Gamma_{\text{Dop}}$ is the Doppler width of Rb atoms ($\sim$500~MHz). The Rabi frequency can be estimated from $\Omega_C = 2\pi\times \Gamma_N \sqrt{I/8}$ \cite{25}, where $I$ is the laser intensity in mW/cm$^2$, and the natural linewidth is $\Gamma_N = 2\pi \times 6$ MHz. In our case, $\Omega_C  = 2\pi \times 7.5$~MHz.

It is known that the spectral width of $\gamma_{\text{EIT}}$ reaches a minimum value when using coherently coupled probe and coupling radiation beams \cite{13,14,15}. However, since our setup uses two coherently uncoupled radiation beams, the width of the EIT$_T$ resonance experiences additional broadening, $\gamma_{\text{add}}$. Therefore, in our case  we add it into Eq. \eqref{eq:eq3}, which leads to $\gamma_{\text{EIT}} = 4.5 \text{ MHz} + \gamma_{c} + \gamma_{\text{add}} = 9 \text{ MHz}$. 
In \cite{24}, a spectral linewidth of $\gamma_{\text{EIT}} \approx 15$ kHz was obtained using low-intensity coherently coupled probe and coupling radiation beams and a $\sim$500~$\mu$m thin cell filled with Cs atomic vapor. Therefore, for simplicity we can assume $\gamma_c$ is not larger than that value and we can conclude $\gamma_{\text{add}} \approx 4.5$ MHz.

Despite the fact that using two coherently uncoupled radiation beams leads to additional broadening of the EIT resonance, the simplicity of forming the EIT resonance with two independent lasers ensures that this method continues to be widely used \cite{26,27,28,29}.

\section{Conclusions}\label{conclusions}

This work demonstrates, for the first time, that it is advantageous to use  selective reflection radiation as a probe field to observe an EIT resonance in small thicknesses of the atomic vapor column ($L \leq 1 \, \mu$m) and low laser powers. An EIT$_{SR}$ resonance is successfully formed in a nanocell with Rb atomic vapor down to thicknesses of $L \approx 150$ nm, using coupling radiation resonant with the corresponding atomic transition. We have compared EIT in SR with EIT observed in the transmitted field and we have shown that for small thicknesses of the atomic vapor column ($L \leq 1 \, \mu$m) EIT with better contrast and narrower linewidth is observed. At thicknesses around 100~nm, EIT$_{SR}$ resonance can be well registered by double differentiation of the initial spectrum, although its spectrum broadens due to frequent collisions of atoms with the windows of the nanocell. 

Conversely, the EIT$_{T}$ resonance formed using micrometric thick cells ($L \geq 50$~$\mu$m) exhibits larger amplitude and narrower spectral width than the EIT$_{SR}$ resonance. This difference is expected, as the SR process operates effectively with an atomic vapor column of $\leq 1000$~nm. Therefore, increasing the cell thickness has a minimal effect on the efficiency of the EIT$_{SR}$ process. In the case of EIT$_{T}$ formation, increasing the cell thickness positively affects the process efficiency, as the number of atoms contributing to the formation of EIT$_{T}$ increases.

Notably, the contrast and spectral linewidth of both EIT$_{SR}$ and EIT$_{T}$ resonances can be improved by using coherently coupled probe and coupling radiation beams \cite{13,14,15}. This approach could potentially allow the formation of EIT$_{SR}$ resonance using SR of atomic vapor column thicknesses as small as 20~nm, for which the SR light is less than 1\% of the incident radiation. The results obtained in the current study should be considered when developing miniature micrometric and nanometric thick optical devices \cite{14}.

The work was supported by the Higher Education and Science Committee of Armenia, in the frames of project N 1-6/IPR.

\printbibliography
\end{document}